\documentclass[a4paper]{article}
\usepackage{graphicx}

\begin{document}

\begin{center}
{\bf \Large
Simulation of ratio of old to young people in countries like Poland}
\bigskip

{\large
D. Stauffer*$^{\dag}$
}
\bigskip

{\em
Faculty of Physics and Applied Computer Science,
AGH University of Science and Technology,
al. Mickiewicza 30, PL-30059 Krak\'ow, Euroland\\
\medskip
* visiting from: Institute of Theoretical Physics, Cologne University,\\
D-50923 K\"oln, Euroland. Visit supported by COST P10.
}

\bigskip
$^\dag${\tt stauffer@thp.uni-koeln.de}, 

\bigskip
\today
\end{center}

\begin{abstract}
\noindent
Countries like Poland with a recent sharp drop in birth rates still have 
some time to prepare for the problems of an ageing society. Ater the
year 2030 they can become increasingly serious.
\end{abstract}

\bigskip
Retirement gets difficult to support if everybody lives longer, the 
birth rates go down, and retirement age and immigration/emigration remain 
constant. This demographic change 
influences the ratio of people in retirement age to those of 
working age. These effects have been simulated by statistical offices in 
many countries, but also with detailed assumptions in journals and a book 
\cite{stauffer,martins,zekri,newbook,sumour}; the latter two references give
a complete Fortran program. 

These methods, with minor adjustments, were used 
to predict the future growth of this ratio for countries with low birth
rates like Germany \cite{stauffer,martins,bomsdorf}, intermediate birth rates
like Algeria, \cite{zekri} and high birth rates like the Palestinian territories
\cite{sumour}, always ignoring special historical events which are particular
for this country. Now we apply this method to a country like Poland where
massive immigration may be less realistic; instead we simulate increases of 
birth rate and retirement age. Simulation details are shifted 
to an appendix.

The decay of the birth rate (more precisely, the average number of children 
per women, unfortunately called the total fertility rate) came in Poland
later but sharper than in Germany and is approximated by 
$2.3-0.55*[1+{\rm tanh(0.15(year}-1993))]$, Fig.1. The total simulated
population, normalized by the actual Polish population in 2002,  agrees
well with what the official Polish authority, www.stat.gov.pl, predicted:
Fig.2. Fig.3 shows the ratio of people above retirement age of 63 
to the people between 20 and retirement age (lower curve). (Our time units
are years throughout.)

\begin{figure}
\begin{center}
\includegraphics[scale=0.45,angle=-90]{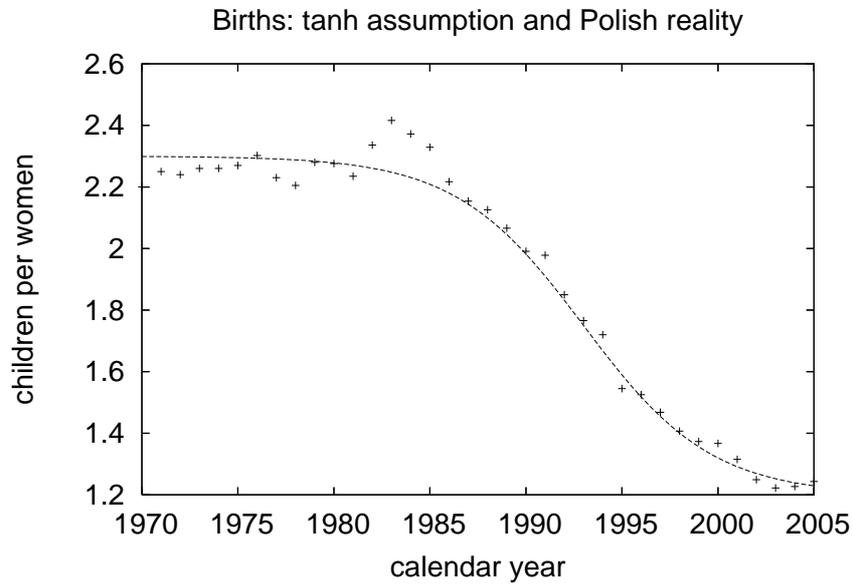}\\
\end{center}
\caption{Real and assumed births. The + come from the official Polish 
statistics, the line is the tannh approximation used in the present 
simulations.
}
\end{figure}

\begin{figure}
\begin{center}
\includegraphics[scale=0.45,angle=-90]{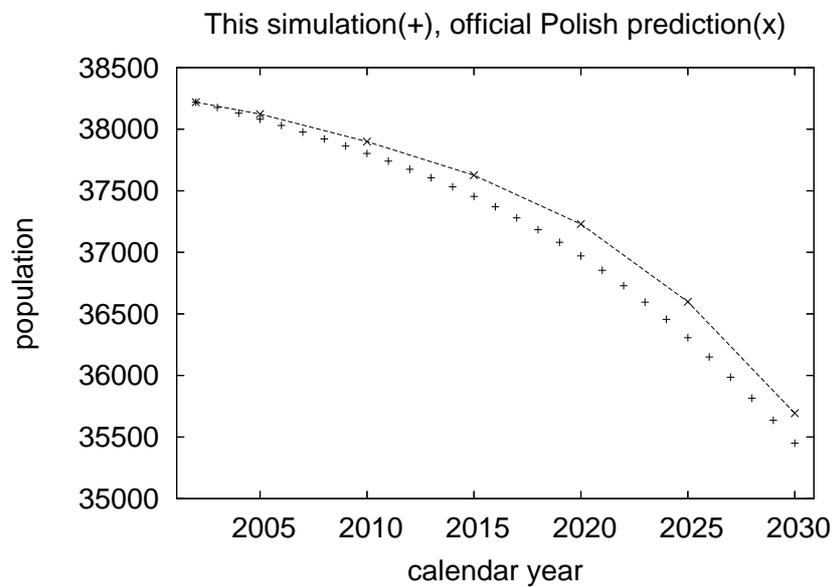}\\
\end{center}
\caption{Official (line) and present (+ symbols) extrapolations of Polish
population.
}
\end{figure}

\begin{figure}
\begin{center}
\includegraphics[scale=0.45,angle=-90]{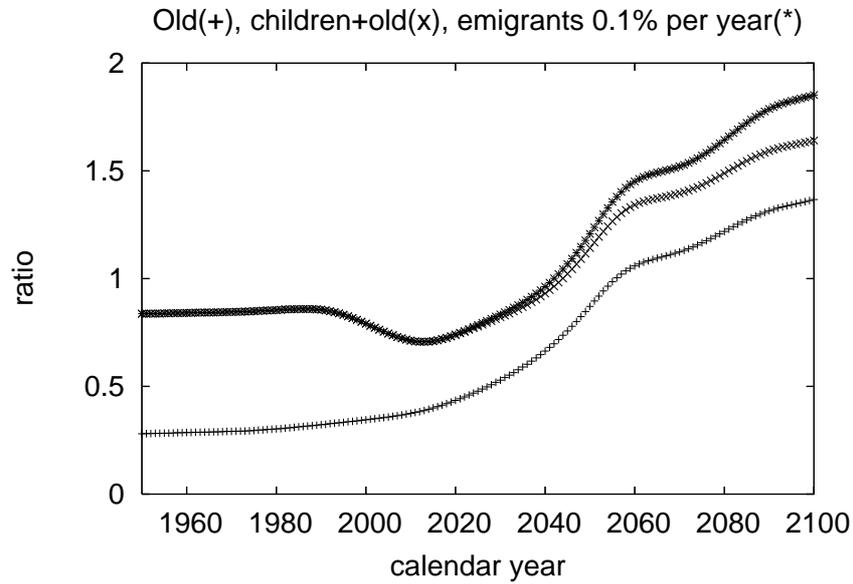}\\
\end{center}
\caption{Ratio of old to working-age people. For the two upper curves the
young people were added to the old ones. Border ages are 20 and 63.
}
\end{figure}

\begin{figure}
\begin{center}
\includegraphics[scale=0.45,angle=-90]{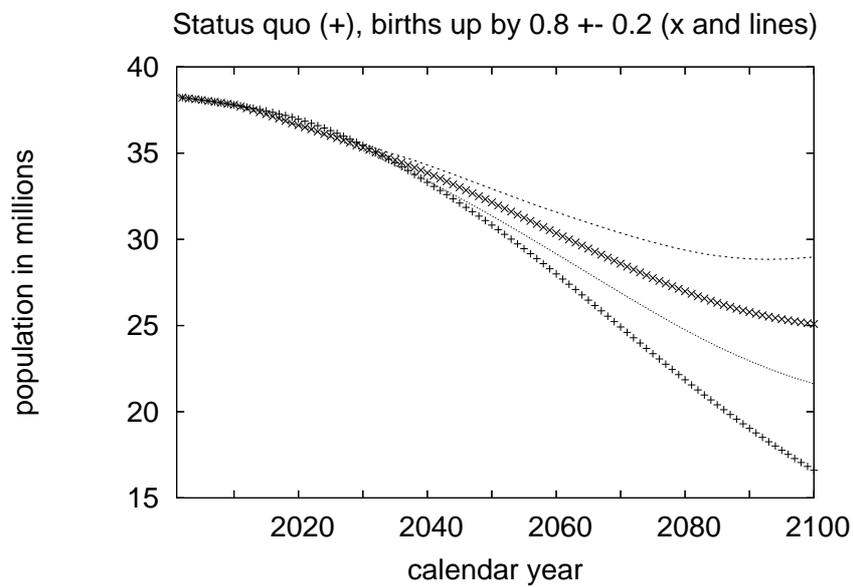}\\
\end{center}
\caption{Help from increased births. The lowest curve uses the present birth
rate, the higher curves assume more or less strong increases of births
in the future.
}
\end{figure}

\begin{figure}
\begin{center}
\includegraphics[scale=0.45,angle=-90]{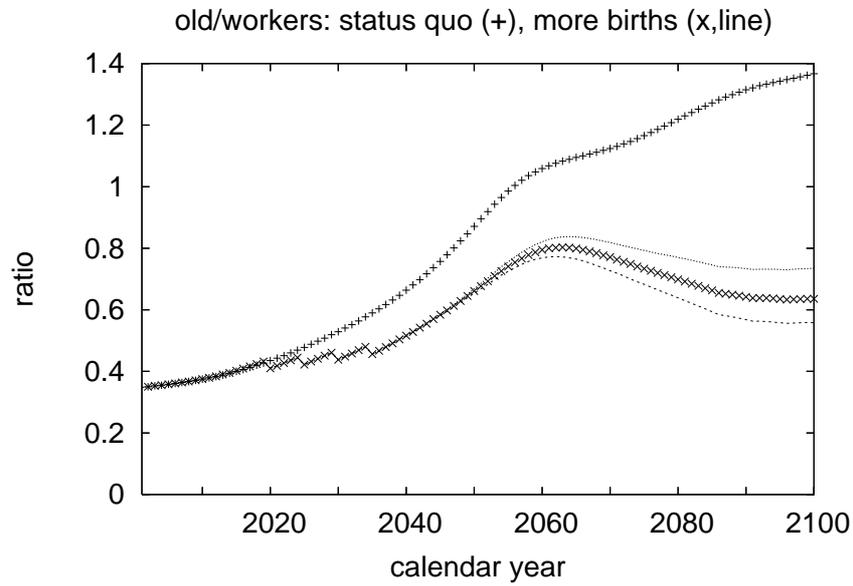}\\
\end{center}
\caption{Help from increased births and increased retirement age. Same 
simulation as in the previous figure; the rise of retirement age from 63 to 67 
does not affect the total population shown in the previous figure.
}
\end{figure}

\begin{figure}
\begin{center}
\includegraphics[scale=0.45,angle=-90]{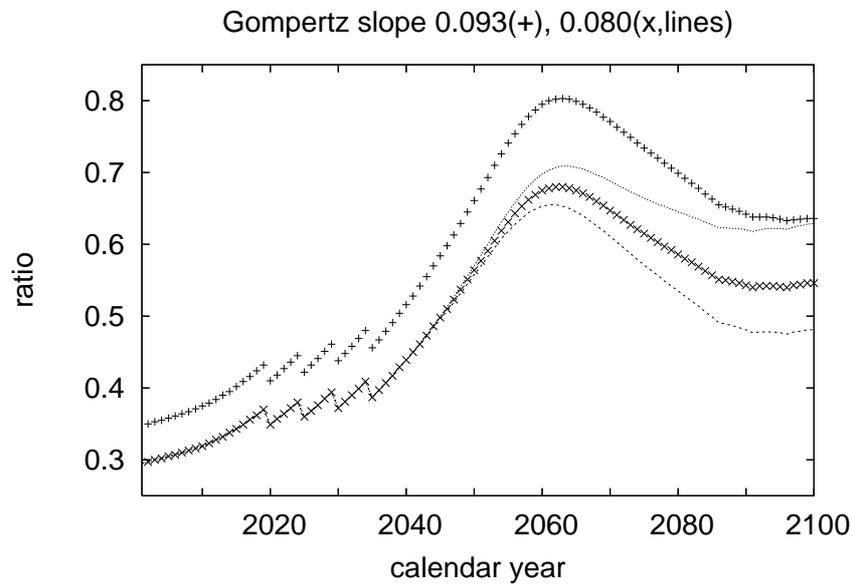}\\
\end{center}
\caption{Ratio of old to working-age people for different Gompertz slopes
0.093 (plus signs, Germany) and 0.08 (other curves, Polish men). 
} 
\end{figure}

\begin{figure}
\begin{center}
\includegraphics[scale=0.45,angle=-90]{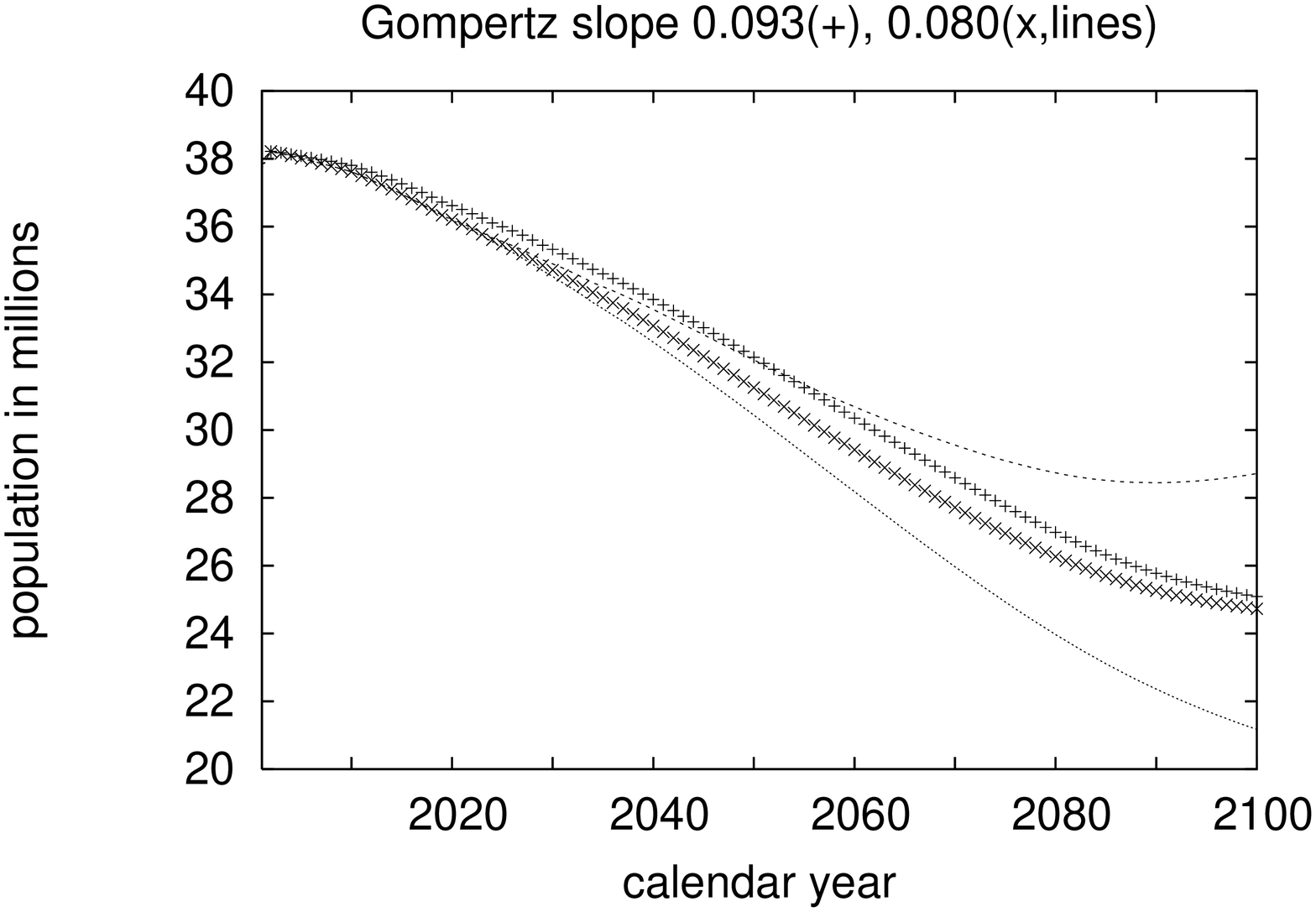}\\
\end{center}
\caption{Populations for different Gompertz slopes, same simulations as in
previous figure.
}
\end{figure}

The future looks less problematic if the number of people up to age 20 is
added to those above 63, both groups needing public support. Then the 
ratio is approaching a minimum: middle curve in Fig.3. Thus for some time,
according to Fig.3 until about 2030, the fraction of people needing support
from the working population will not be much higher than it was in the
past. THus one has some years time to think, discuss and agree on how
to solve the future problems of demographic change. The upper curve
there shows the effects of a net emigration of 0.1 percent per year,
starting in 2010 (e.g. from Poland to Western Europe; some statistical
data give higher values already now). In the past
this emigration was negligibly small, about 0.04 percent per year. 
The lowest curve in Fig.4 shows the resulting decay of the total 
population.

To get more optimistic curves, the retirement age was assumed to increase 
from 63 by one year and the births by 0.2 in the years 2020, 2025, 2030 and 
2035; increase in retirement age alone did not help much. These assumptions
give the x symbols in Fig.4. The two lines on both sides of the x curve
correspond to changes 0.25 (higher line) and 0.15 (lower line) in the
births and indicate the order of magnitude of the extrapolation errors. 
Fig.5 shows the ratios of people in retirement age to people in working age,
corresponding to the same simulations as in Fig.4. 

In the above simulations the Gompertz slope was taken as $b=0.093$, with
the mortality function increasing for adults as $\exp(bx)$ with increasing
age $x$, as in \cite{stauffer,martins}. The actual Polish value is near 0.08
similar to Algeria\cite{zekri}. Using $b = 0.08$ instead of 0.093 we get Figs.6
and 7, which overlap with the results of Figs.4 and 5. 

Thus enhanced birth rates as simulated here and massive immigration as simulated
for countries like Germany \cite{stauffer,newbook} could reduce the shrinking 
of the population and the burden of the working population to support the 
retired people. That burden would also be alleviated by increasing retirement
age; German parliament adopted in 2006 a law regulating these future increases.
France recently increased the births from 1.7 to 1.9 within a decade. Other
countries in the European Union, like Bulgaria, Romania or the three Baltic 
states, may be in a situation similar to Poland. 

Simulations like these took less than a second each, in contrast to more
sophisticated methods \cite{bonkowska}, and readers can
change parameters to check for the effects of different assumptions. 
Countries corresponding to these simulations still have time to adjust
to the future problems of the demographic change.
The ageing problems seem to become very serious after the year 2050.

K. Ku{\l}akowski kindly provided the empirical Polish data from 
www.stat.gov.pl used in this work, and commented on the manuscript. 

\bigskip

{\centerline \bf Appendix:}

\bigskip
The simulations calculate the age distribution of the population in one year
from that in the preceding year, neglecting possible correlations between
a mother and her daughters. For extrapolations over a few generations
this approximation should be good enough. A complete Fortran program is given
in \cite{newbook,sumour}.

Women give birth from ages 21 and 40, of daughters with half of the birth
number given in the text, spread evenly over these 20 years. Sons 
represent the other half and can be neglected. The population increases by 
immigration or decreases by emigration, affecting equally all ages from 6 to 40. 
The mortality is assumed as $7b\ \exp[b(x-X)$, giving a survival probability
$S(x)$ from birth to age $x$ as $S = \exp(-7 \exp(-bX)(\exp(bx)-1))$. Thus
after births and migration have been dealt with, the population $P(x)$ at
age $x$ is calculated from $P(x) = P(x-1) S(x)/S(x-1)$. Here the Gompertz
slope $b$ is assumed to increase over 150 years from $b = 0.07$ to $b = 0.093$ 
(or 0.08) until the year 1971, and then to stay constant at this maximum value. 
The characteristic age $X$, in contrast, is assumed to stay constant at 103
until 1971, and thereafter to increase by 0.15 each year. This change of
trends around 1971 was seen in some empirical studies of the last years.

\end{document}